\documentclass[preprint,trackchanges,times,longauthor]{aastex61}
\hypersetup{linkcolor=red,citecolor=blue,filecolor=cyan,urlcolor=magenta}

\usepackage{natbib}
\usepackage{amsmath}
\usepackage{graphicx}
\usepackage{float}
\usepackage{rotating}
\usepackage[toc,page]{appendix}
\usepackage{epsfig}
\submitjournal{ApJ}
\shorttitle{Alfv\'{e}n wave embedded magnetic cloud and  geomagnetic storm }
\shortauthors{Raghav et al.}

\begin{document}

\title{Torsional Alfv\'{e}n wave embedded ICME magnetic cloud and corresponding geomagnetic storm. }

\correspondingauthor{Anil Raghav}
\email{raghavanil1984@gmail.com}

\author[0000-0002-4704-6706]{Anil N. Raghav}
\affil{University Department of Physics, University of Mumbai, Vidyanagari, Santacruz (E), Mumbai-400098, India}

\author{Ankita Kule}
\affiliation{University Department of Physics, University of Mumbai, Vidyanagari, Santacruz (E), Mumbai-400098, India}
\author{Ankush Bhaskar}
\affiliation{Heliophysics Science Division, NASA/Goddard Space Flight Center, Greenbelt, USA}
\affiliation{University Corporation for Atmospheric Research, Boulder, USA}
\author{Wageesh Mishra}
\affiliation{University of Science and Technology of China, China}

\author{Geeta Vichare}
\affiliation{Indian Institute of Geomagnetism, Navi Mumbai India.}

\author{Shobha Surve}
\affiliation{Centre for Excellence in Basic Sciences, Mumbai, India}

\begin{abstract}
The energy transfer during interaction of large-scale solar wind structure and the Earth's magnetosphere is the chronic issue in space-weather studies. To understand this, researchers widely studied the geomagnetic storms and sub-storms phenomena. The present understanding suggests that long duration of southward interplanetary magnetic field component is the most important parameter for the geomagnetic storm. Such long duration strong southward magnetic field is often associated with ICMEs, torsional Alfv\'{e}n fluctuations superposed co-rotating interacting regions (CIRs) and fast solar wind streams. Torsional Alfv\'{e}n fluctuations embedded CIRs have been known for a long, however magnetic cloud embedded with such fluctuations are rarely  observed. The presence of Alfv\'{e}n waves in the ICME/MC and influence of these waves on the storm evolution remains an interesting topic of study. The present work confirms the torsional Alfv\'{e}n waves in a magnetic cloud associated with a CME launched on 15th February  which impacted the Earth's magnetosphere on February 18, 2011. Further, observations indicate that these waves inject energy into the magnetosphere during the storm and contribute to the long recovery time of geomagnetic storms. Our study suggests that presence of torsional Alfv\'{e}n waves significantly controls the storm dynamics.

\end{abstract}


\section{Introduction} \label{sec:intro}
In the present satellite era, the study of space physics has become of paramount importance as space weather events can disturb our day-to-day activities. Geomagnetic storms \textit{i.e.} the temporary disturbances in the Earth's magnetosphere \citep{lakhina2007humboldt,gonzalez1994geomagnetic,tsurutani1995interplanetary}, are one of such events that can lead to serious adverse effects on communication, navigation and electrical systems \citep{carrington1859description,cannon2013extreme,schrijver2010heliophysics,osella1998currents,tsurutani2003extreme,lakhina2016geomagnetic,gawali2016science}. In general, the southward oriented interplanetary magnetic field (IMF) in combination with a strong, high speed solar wind stream are considered as a main driver of a geomagnetic storm  \citep{gonzalez1999interplanetary,tsurutani1992great,tsurutani1995interplanetary,vichare2005some}. Whereas a viscous type of interactions of high speed solar-wind with magnetosphere are also considered when IMF is northward \citep{axford1964viscous}. However, the energy transferred to the magnetosphere through this is very small and therefore, it is no longer believed to be geo-effective for creating geomagnetic storms \citep{tsurutani1995efficiency,du2008anomalous}. The negative $B_z$ component interacts with Earth's magnetic field via magnetic reconnection and the charged particles penetrate into the Earth's magnetosphere from day-side of the magnetosphere \citep{gonzalez1994geomagnetic}. Most of the particles are injected into the inner magnetosphere because of plasma sheet reconnection at the magnetotail \citep{behannon1966magnetic}.  Due to the gradient and curvature of the Earth's magnetic field, injected particles \textit{e.g.} protons drift towards west and electron towards the east such that westward ring current is set up. The magnetic moment of the ring current is in the opposite direction as that of the Earth's magnetic dipole moment. Thus, a negative magnetic field perturbation \textit{i.e.} a depression is observed at the ground.


The literature on the geomagnetic effects of solar wind structures identify three major causes of geomagnetic storms; (i) Coronal Mass Ejections (CMEs); (2) Co-rotating interaction regions (CIRs); and (3) the Alfv\'{e}n waves superposed high speed solar wind streams \citep{tsurutani1997interplanetary,tsurutani2006corotating,cane1997caused,parker1958interaction,tsurutani1988origin,borovsky2006differences}. The CME-driven storms are frequently stronger with significantly enhanced ring currents which induce great auroras. The CIR-driven storms are normally moderate/weaker with less intensive auroras \citep{borovsky2006differences,tsurutani1997interplanetary}.
The geomagnetic storm comprises various phases viz. sudden storm commencement(SSC) or gradual commencement (GC), initial phase, main phase and recovery phase of the storm \citep{tsurutani1997interplanetary,gonzalez1999interplanetary}. The SSC and/or GC are mainly controlled by the solar wind dynamic pressure i.e. initiated by forward shock-front of CME. 
In CME-driven storms, the main phase cause by a southward IMF which directly contribute to increase in ring current whereas the recovery start when ring currents begin to decay. Thus CME-driven storms only last for few days \citep{borovsky2006differences}. 

As compare to this, the CIR and High Speed Stream (HSS) from coronal holes driven magnetic storms can last for a few days up to few weeks \citep{tsurutani2006corotating}, generally known as High Intensity Long Duration Continuous AE Activity (HILDCAAs) events \citep{tsurutani1987cause,hajra2013solar,hajra2014relativistic,hajra2015relativistic,prestes2017high}. The CIR occurs at the interface between the HSS and the upstream slow stream \citep{smith1976observations,burlaga1977causes,gonzalez1999interplanetary}. The amplified Alfv\'{e}nic fluctuations are observed within CIRs as well as the HSS that trail the CIR (has lower amplitude Alfv\'{e}n waves). Magnetic reconnection between the southward component of the Alfv\'{e}n waves and the magnetosphere fields slowly inject solar wind  energy and plasma into the magnetosphere which cause the extended recoveries of the storms \citep{tsurutani2006corotating,tsurutani1995large}.


Although Alfv\'{e}n waves are very common in space plasmas, it is rare to obtain their spatial picture. This is because Alfv\'{e}n waves are slowly varying and have very long wavelengths along the magnetic field. Due to torsional wave, there are no fluctuations in the fluid density hence no variations in the observed emission intensity images and thus they are more difficult to be observed. However, based on the spectrometric studies of solar corona, the presence of Alfv\'{e}n wave is inferred \citep{banerjee1998broadening, banerjee2009signatures,harrison2002off}. Recently \citep{tian2010signatures} proposed that Alfv\'{e}nic fluctuations are observed in flux ropes. \citep{gosling2010torsional} also suggested that a torsional wave could be generated by distortions within a flux rope ejected from the Sun. It is considered that (i) twisting in straight or homogeneous magnetic field lines, (ii) ejecting axial current at the base of pre-existing flux rope, or (iii) otherwise changing the force balance condition associated  with pre-existing flux rope can generate Alfv\'{e}n waves   \citep{gosling2010torsional,fan2009emergence}. Therefore, Alfv\'{e}n waves are expected to be observed within magnetic cloud like structures. However, they are not commonly observed within the magnetic clouds of various sizes in the solar wind. In earlier studies, Alfv\'{e}nic wave fluctuations are observed in the region where fast and slow solar wind streams interact \textit{i.e} CIRs region \citep{tsurutani1995large,lepping1997wind}. Further, the presence of the Alfv\'{e}n waves during interface of magnetic cloud and solar wind stream is observed  \citep{lepping1997wind,behannon1991structure}. Recently, the presence of Alfv\'{e}n waves embedded magnetic cloud is interpreted as the output of CME-CME interaction \citep{raghav2017energy}. In the present work, we aim to study the presence of torsional Alfven waves in magnetic cloud (MC) and trailing region of the MC. It is not so common to observe  Alfv\'{e}n waves in magnetic cloud. This motivates us to understand the geomagnetic storm characteristics associated with both aforementioned regions. Finally we discuss the possible contribution of Alfv\'{e}n waves in the storm and compare with past studies.  

\section{Observations}

The event under study caused by the multiple-CME’s erupted on $13^{th} $, $14^{th}$ and $15^{th}$ February 2011, interacted en route and appeared as complex magnetic structure at 1 AU in WIND satellite data. The selected event has been studied in past to understand: 1) their interaction corresponding to different position angles \citep{temmer2014asymmetry}, 2) their geometrical properties and the coefficient of restitution for the head-on collision scenario and their geomagnetic response \citep{mishra2014morphological}, 3) corresponding Forbush decrease phenomena  \citep{raghav2014quantitative,marivcic2014kinematics,raghav2017forbush}, 4. the presence of Alfv\'{e}n waves in interacting region \citep{raghav2017energy}. 

\begin{figure}
\includegraphics[width=0.8\textwidth]{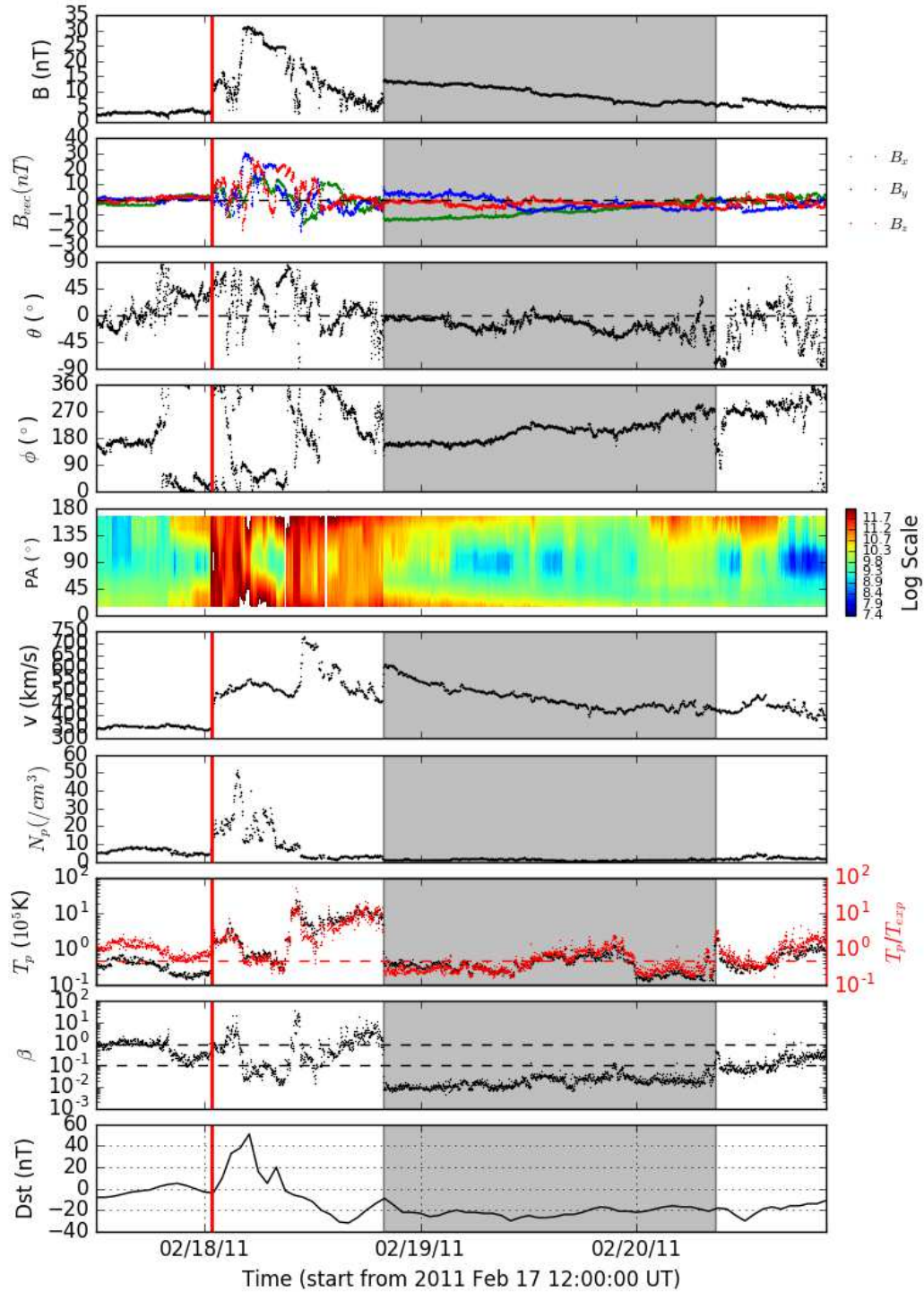}
\caption{The interplanetary magnetic field (IMF), solar wind plasmas and the suprathermal electron pitch angle distribution observations of complex CME-CME interaction event crossed on 18-19 February 2011 from WIND spacecraft. From top to the bottom, parameters are the magnetic field strength ($|B|$), the vector components of the magnetic field in GSE coordinate system, the elevation (theta) and azimuthal (phi) of field direction in GSE coordinate system, the suprathermal electron pitch angle distribution, solar wind speed ($V_{SW}$), proton density ($N_p$), proton temperature ($T_p$), the ratio of proton thermal pressure to magnetic pressure ($\beta$) and the $D_{st}$ index from WDC. The red dashed lines denote the arrival time of the shock. The gray shades show the region of the ICMEs magnetic cloud. Courtesy of figure  \url{http://space.ustc.edu.cn/dreams/wind_icmes/index.php} }
\label{fig:1}
\end{figure}

The Figure \ref{fig:1} shows the in situ observation (in GSE- coordinate system) of the ICMEs with $D_{st}$ index. The red lines depict the arrival time of the shock and the gray shades show the region of the ICMEs magnetic cloud. The top panel displays magnitude of the magnetic field $B~(nT)$ which shows magnetic field enhancement and gradual decrease in shaded region. The second top panel consists of magnetic vectors $B_x$, $B_y$ and $B_z$, in which $B_z$ shows no rotation, $B_y$ shows slight rotation whereas $B_x$ shows smooth variations in shaded region. The third panel from top consists of  $\theta$ angle which is relatively constant and the fourth-panel consist of $\phi$ angle varies slowly in shaded region.  Hence, this suggests a magnetic cloud or magnetic cloud-like structure with relatively small rotation. It is also identified as magnetic cloud in the Lepping MC list available at  \url{https://wind.gsfc.nasa.gov/mfi/mag_cloud_S1.html}. The fifth panel shows pitch angle distribution of electrons which indicate signature of MC. The sixth panel display gradually decreasing solar wind speed. The seventh panel demonstrate almost constant low proton density (Np). The eight panel consists of proton temperature on left side and ratio of proton temperature observed to expected proton temperature on right which is nearly $0.5$ through out the MC crossing. The second panel from bottom display less than $0.1$ value of plasma beta within the MC boundaries. The bottom panel consists of $D_{st}$ index which suggest overall the geomagnetic response to observed MC. The long recovery phase with moderate/weak type of storm is observed during the observed MC.

\begin{figure}
\includegraphics[width=1 \textwidth]{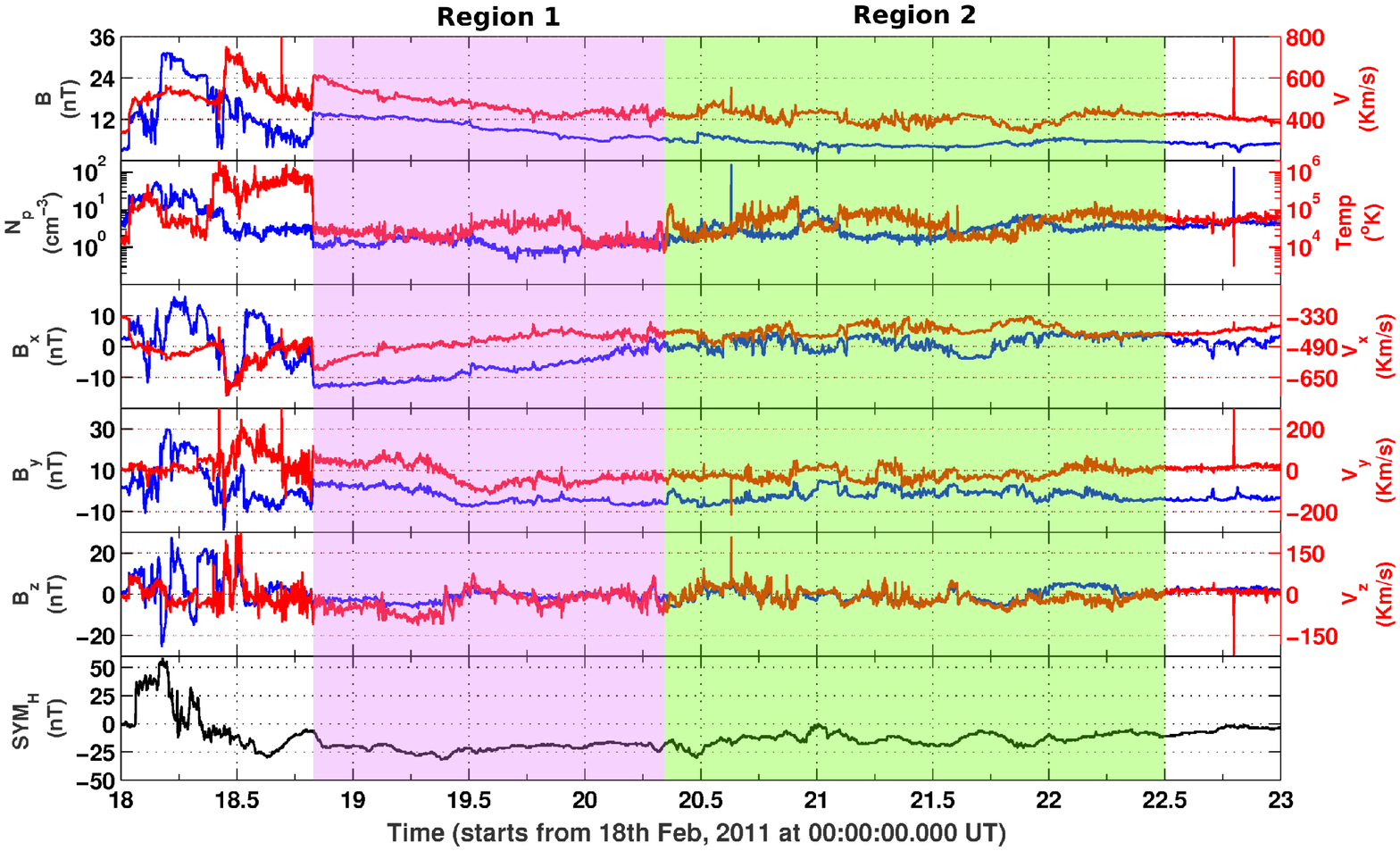}
\caption{From top to the bottom, the first panel shows the magnetic field strength ($|B|$) and solar wind speed ($V_{SW}$), second panel shows proton density ($N_p$), proton temperature ($T_p$), the 3rd, 4th, and 5th panel show the vector components of the magnetic field and solar wind velocity  in GSE coordinate system respectively (data taken from Wind satellite with time cadence of 92 sec). The bottom panel show the $SYM_H$ profile (data taken from Omni database with 60 sec time resolution). The pink shaded region is the MC like structure and green shaded region is the trailing solar wind after CME.}
\label{fig:2}
\end{figure}

In Figure \ref{fig:2}, the region 1 and region 2 is specifically shown with color shade of pink and green. The observed complex interplanetary structures before region 1 is explained at length in  \citep{marivcic2014kinematics,mishra2014morphological,raghav2017energy}. Region 1 is shaded with pink color and it is identified as MC/ MC-like structure. Similarly, Region 2 is (green shade color) MC trailing solar wind. The third, forth and fifth panel from top demonstrate the temporal variation of magnetic field and solar wind components in GSE-coordinate system. The peculiar and distinct feature of variation in all three components of $B$ and $V$ is seen in region 1 and 2 i.e. 
the well-correlated changes in respective components of magnetic field $B$ and solar wind speed $V$ is observed.  \cite{raghav2017energy} manifest this as presence of Alfv\'{e}n waves in MC i.e. region 1. Further, the region 2 also shows similar trend of variations which might be Alfv\'{e}nic type. The bottom panel display the temporal variations of longitudinal symmetric disturbance index SYM-H. It remains almost constant during complete passage of MC and suggest weak/moderate disturbed condition with approximately $-25~ nT$ value. After MC cross over the SYM-H recovers to $\sim$ $0~ nT$ at $00:00~UT$ on 21 February and further remains fluctuating in negative values in region 2. 

\begin{figure}[ht]
\includegraphics[width=1 \textwidth]{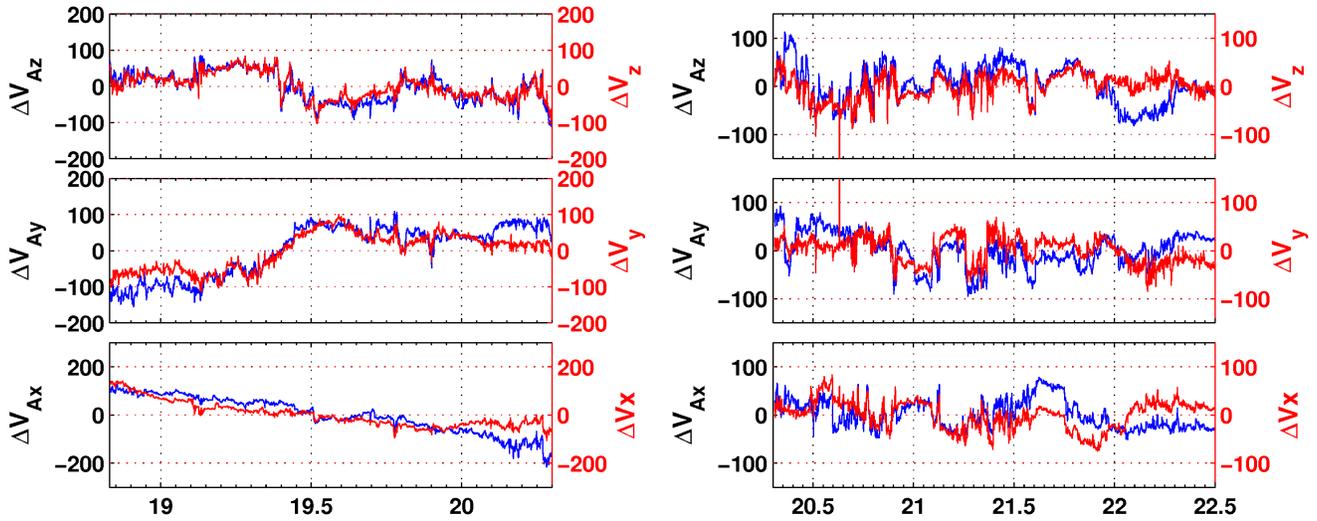}
\caption{Right 3 panels (region 1) and left 3 panels (region 2) illustrate relative fluctuation of Alfv\'{e}n velocity vector $\Delta V_A$ (blue lines) and that of proton flow velocity vector $\Delta V$ (red lines). ( time cadence of 92 sec)}
\label{fig:4}
\end{figure}

To confirm Alfv\'{e}n waves presence in region 1 and possibility in region 2 the Wal\'{e}n relation is used. The Wal\'{e}n relation is described as
\begin{equation}
  V_A = \pm A~ \frac{B}{\sqrt{\mu_0 \rho}}
\end{equation} 
where $A$ is the anisotropy parameter, $B$ is magnetic field vector and $\rho$ is proton mass density \citep{walen1944theory,hudson1971rotational}. By considering negligible influence of the thermal anisotropy one can ignored anisotropy in the solar wind, therefore,
we usually take $A=\pm1$ \citep{yang2016observational}.
The fluctuations $\Delta B$ in $B$ can be obtained by subtracting average value of $B$ from each measured values. Therefore, the fluctuations in Alfv\'{e}n velocity is 
\begin{equation}
\Delta V_A = \frac{\Delta B}{\sqrt{\mu_0 \rho}}
\end{equation}
Furthermore, the fluctuations of proton flow velocity $\Delta V$ is estimated by subtracting averaged proton flow velocity from measured values. Figure \ref{fig:4}, in top, middle and bottom  panels shows the comparisons of x, y and z components of $\Delta V_A$ and $\Delta V$, respectively. The fluctuations seen in Figure \ref{fig:4} for both the regions indicate the presence signature of Alfv\'{e}n waves. However, the visual variations in region 1 are appeared to be strongly correlated as compared to region 2.

\begin{figure}[ht]
\includegraphics[width=1 \textwidth]{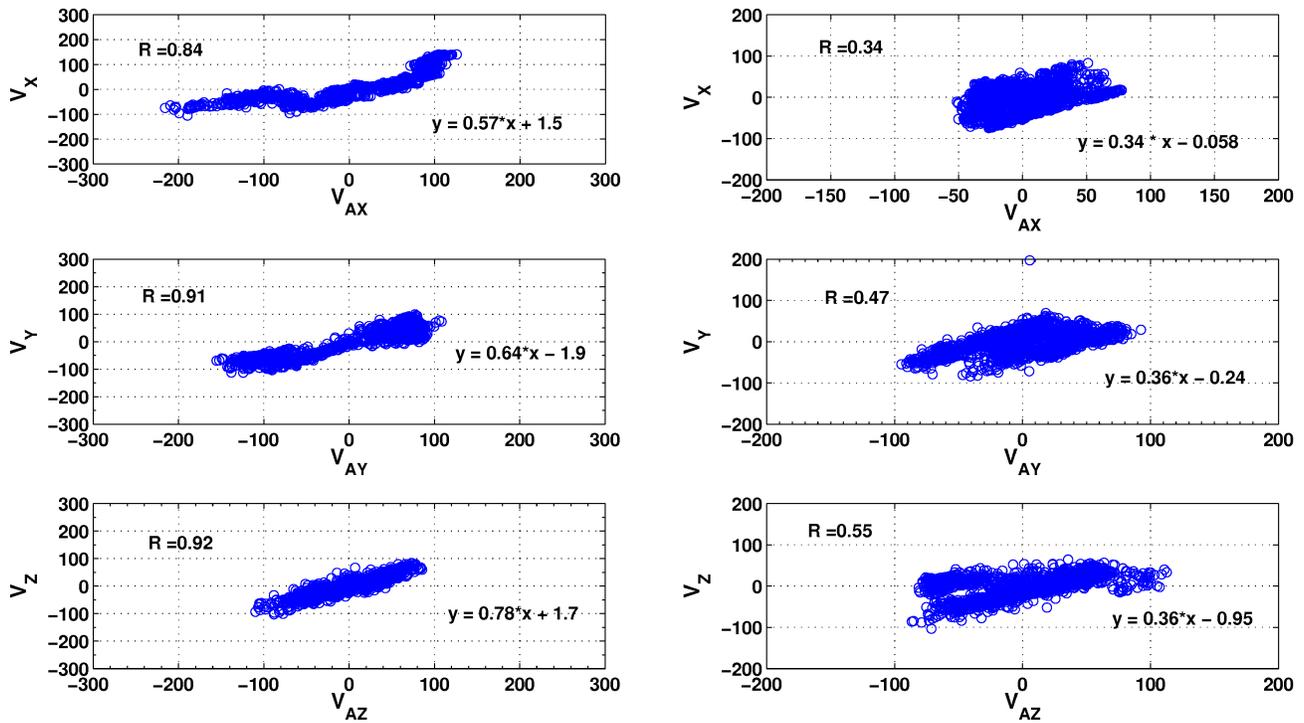}
\caption{The linear relation between $\Delta V_A$ and $\Delta V$ for region 1 (left) and region 2 (right) for the event shown in Figure \ref{fig:2}.   The scattered blue circles are observations from Wind satellite with time cadence of 92 sec. The R is the correlation coefficient. The equation in each panel suggest the straight-line fit relation between respective components of $\Delta V_A$ and $\Delta V$.}
\label{fig:5}
\end{figure}

Moreover, the Figure \ref{fig:5} shows the correlation and linear relation between the fluctuations of Alfv\'{e}n velocity vector components and the fluctuations of proton flow velocity vector components for both the region 1 and region 2 as marked in Figure \ref{fig:2}. The linear equations and correlation coefficients are shown in each panel. For the region 1,  The slopes for x, y and z components of three vectors are 0.57, 0.64 and 0.78 and the correlation coefficient ($R$) between Alfv\'{e}n and proton flow velocity vector components are noted as 0.84, 0.91, and 0.92, respectively. For the region 2,  the slopes for three vectors components are 0.34, 0.36 and 0.36 and the correlation coefficient between Alfv\'{e}n and proton flow velocity  vector components are 0.34, 0.47, and 0.55, respectively.  
 The correlation coefficients suggest strong positive correlation for region 1 which confirms the Alfv\'{e}n waves in MC and weak positive correlation for region 2 between $\Delta V_{Ax}$ \& $\Delta V_x$, $\Delta V_{Ay}$ \& $\Delta V_y$, $\Delta V_{Az}$ \& $\Delta V_z$. The MHD static equilibrium suggests that the MC cloud is considered to the strongly coupled plasma under frozen in condition with concentric cylindrical surface layers around the central axis.
 Therefore the fluctuations in each layer is mutually connected which leads to the strong correlation coefficient. However, in trailing solar wind, the heliospheric current sheets are not strongly coupled. Thus, transfer of fluctuations from one sheet to other are not mutually correlated. This leads to week correlation coefficient in region 2. In fact, different correlated layers are visible in region 2 of Figure ~\ref{fig:5}. The detail study of Alfv\'{e}n fluctuations transfer is needed.
 

\section{Results and Discussion}
From our analysis, we find a good correlation between magnetic field and proton velocity vectors. This suggests that magnetic field and fluids are oscillating together in the observed region 1 of the magnetic cloud. Such a scenario means that fluid velocity perturbation is perpendicular to magnetic tension force which is along the direction of resultant magnetic field. Such a velocity perturbation would lead to magnetic field perturbation and can generate a wave propagating along the direction of tension force. Thus, the wave propagation direction is along the resultant magnetic field direction. This characteristic of the wave is for torsional Alfv\'{e}n wave. Thus, our analysis confirms the presence of Alfv\'{e}n wave in the magnetic cloud and its weak signature in the trailing solar wind flow.  

Earlier studies have shown that HILDCAA events, following a geomagnetic storm, are induced by interplanetary Alfv\'{e}n waves propagating outward with solar wind.  \citep{tsurutani1987cause,prestes2017high}. The presence of co-rotating streams containing continuous, large amplitude Alfv\'{e}n waves causing substorm activity tends to long recovery time of several days \citep{tsurutani1995interplanetary}. The associated RED (Relativistic electron events) differentiate between CME associated storm and Co-rotating streams associated storm depending on the parameters mainly $D_{st}$ index, $B_z$ component, and the solar wind pressure. It was noted that the recovery time for both events was long and $B_z$ component was fluctuating around zero \citep{alyana2007differences}. It is considered that the Alfv\'{e}n waves may lead to long duration southward magnetic field and help in to provide the energy and plasma injection to the magnetosphere. However, the Poynting flux observations of DSMP satellite failed to suggest Alfv\'{e}n waves as an energy input to the magnetic storm and ascribed that the energy input is due to heating of ions and neutral atoms \citep{huang2017dmsp}. \citet{hui1992electron} have demonstrated that electrons can be accelerated by kinetic Alfv\'{e}n waves in both hot and cold plasma.  The long duration and large amplitude Alfv\'{e}n waves have potential to induce different characteristics/profile of geomagnetic storms than that from CMEs and CIRs in the absence of such waves. Thus, Alfv\'{e}n waves play an important role in the dynamics of space plasma from solar corona to 1 AU, and even beyond.

In general, the recovery phase is observed once southward/negative IMF $B_z$ component turns northward/positive \citep{adebesin2009study}. The charge exchange, coulomb scattering and wave-particle interaction are the possible physical mechanisms responsible to the recovery phase \citep{gonzalez1994geomagnetic,daglis1999terrestrial}.
In the charge exchange process, energetic $H^+$ or $O^+$ or $N^+$ ions collide with atmospheric neutral atoms, the ions capture the electrons and become neutralized. The lifetime for charge exchange process depends upon neutral atomic hydrogen or oxygen density and the equatorial pitch angle of ions. Charge exchange process is important when there are singly charged ions($H^+$ or $O^+$ or $N^+$) with energies up to few KeV whereas for higher energies coulomb collision is important mechanism especially for heavier ions  \citep{daglis1999terrestrial}. When the recovery time is slow, then wave-particle interaction could be contributing factor. It is suggested that the pitch angle diffusion by plasma waves also contributes to ring current losses \citep {daglis1999terrestrial}. 
Moreover, it is suspected that during extreme  geomagnetic  storms, the wave  intensities  (both proton electromagnetic  ion  cyclotron (EMIC) waves and electron electromagnetic  (chorus) waves) could  be  substantially  greater resulting in strong wave-particle interaction. Thus, due to the greater loss  cone  size and  the enhanced  wave intensities  rapid ring  current  losses can  occur.  \citep{tsurutani2009comment}. Therefore, wave coherency \citep{tsurutani2009properties,lakhina2010pitch,bellan2013pitch,remya2015electromagnetic} in  wave-particle  cyclotron  resonant  interactions should  be  considered in any updated model explaining magnetic storm \citep{tsurutani2009comment}.

Here, we showed that a unusual long recovery time  is observed for minor geomagnetic storm which was caused by MC of ICME. It is important to note that the $B_z$ component of IMF remained fluctuating near zero through out the region 1 and 2 (see figure \ref{fig:1} and \ref{fig:2}). We interpret that long duration of $B_z$  component fluctuations near zero value is maintained due to the presence of embedded Alfv\'{e}n waves. Each planetary magnetosphere is considered as one type of plasmoid in the heliosphere, with a particular orientation of magnetic field. Our earlier study suggested that CME-CME interaction may give rise to magneto-hydrodynamic ( torsional Alfv\'{e}n) waves in ICMEs or magnetic clouds  which implied to CME-magnetosphere interaction as well \citep{raghav2017energy}. In a condition of oppositely magnetic orientation of the plasmoids, the magnetic reconnection is the  possible physical mechanism for ICME and palsmoid interaction. Thus the present study suggests that the fluctuating $B_z$ fields comprising Alfv\'{e}n wave are expected to prolong the ring current decay by injecting solar wind energy into the magnetosphere continuously. This is similar to the High-Intensity, Long Duration, Continuous AE Activity (HILDCAAs)\citep{tsurutani1987cause} but caused by magnetic clouds. This indicate the presence of torsional Alfv\'{e}n waves may contribute substantially to geo-effectiveness of solar wind structures even magnetic cloud.   \citep{zhang2014alfven}.

The present study demonstrated the presence of Alfv\'{e}n waves in a magnetic cloud and trailing solar wind flow associated with the CME of 15th February. This CME have interacted with preceding CMEs ejected on 13th and 14th February during its heliospheric evolution. \citet{raghav2017energy} suggested that their interaction may form merged interacting complex ejecta structure and change the force balance conditions of flux ropes which leads to generate the Alfv\'{e}n wave. It is possible that interaction of CMEs causes a reconnection between field lines and Alfv\'{e}n waves may be generated at such reconnection point. It is  also possible that with the steepening of a magnetosonic wave which forms the shock at the leading edge could create a trailing Alfv\'{e}n wave \citep{tsurutani1988origin,tsurutani2011review}. Besides this, it has been suggested that Alfv\'{e}n waves may be generated in the interplanetary space \citep{hellinger2008oblique} and they have been observed in the solar wind propagating outward from the Sun. However in that case it would be difficult for the Alfv\'{e}n wave to get into the MC. Further studies are required to understand the distribution of Alfv\'{e}n waves and turbulence features in the magnetic clouds. 

We conclude that the torsional Alfv\'{e}n waves play significant role during the interaction of CME and planetary magnetosphere, they may significantly contribute in geo-effectiveness and recovery time of geomagnetic storms. More studies are required to confirm whether or not a majority of  magnetic clouds embedded with Alfv\'{e}n waves lengthen the recovery phase of geomagnetic storms.  Also, this suggest that one needs to consider  energy injection by Alfv\'{e}n waves during the recovery phase in addition to other loss processes like charge exchange, wave-particle interactions etc.  Further studies are required to investigate if the Alfv\'{e}n waves in ICMEs/MC and within CIRs have different characteristics, possibly because of different plasma beta values in different solar wind structures. And, how the geo-effectiveness is different for them.

  \section{Acknowledgement}
  We are thankful to WIND Spacecraft data providers (wind.nasa.gov) for  making interplanetary data available. We are also thankful  to Department of Physics (Autonomous), University of  Mumbai, for providing us facilities for fulfillment of this work. AR also thanks to Solar-TErrestrial Physics (STEP) group, USTC, china \& SCOSTEP visiting Scholar program. W.M. is supported by the Chinese Academy of Sciences (CAS) President’s International Fellowship Initiative (PIFI) grant No. 2015PE015 and National Natural Science Foundation of China (NSFC) grant No. 41750110481. A.B. is supported by the NASA Living With a Star Jack Eddy Postdoctoral Fellowship Program, administered by UCAR’s Cooperative Programs for the Advancement of Earth System Science (CPAESS)."

\bibliographystyle{aasjournal}

\end{document}